\documentclass[a4paper,showpacs,aps,preprint]{revtex4}%
\usepackage{amsfonts}
\usepackage{amsmath}
\usepackage{amssymb}
\usepackage{graphicx}%
\begin{document}
\preprint{ }
\title[Alpha-Helix Formation]{Dynamics of Alpha-Helix Formation in the CSAW Model}
\author{Jinzhi Lei}
\affiliation{Zhou Pei-Yuan Center for Applied Mathematics, Tsinghua University, Beijing,
China 100084 (Email: jzlei@tsinghua.edu.cn) \ }
\author{Kerson Huang}
\affiliation{Physics Department, Massachusetts Institute of Technology, Cambridge, MA, USA
02139 (Email: kerson@mit.edu)}

\begin{abstract}
We study the folding dynamics of polyalanine (Ala$_{20}$), a protein fragment with 20
residues whose native state is a single alpha helix. We use the CSAW model
(conditioned self-avoiding walk), which treats the protein molecule as a chain
in Brownian motion, with interactions that include hydrophobic forces and
internal hydrogen bonding. We find that large scale structures form before
small scale structures, and obtain the relevant relaxation times. We find
that helix nucleation occurs at two separate points on the protein chain.
The evolution of small and large scale structures involve different
mechanisms. While the former can be describe by rate equations governing the
growth of helical content, the latter is akin to the relaxation of an elastic solid.

\end{abstract}
\volumeyear{year}
\volumenumber{number}
\issuenumber{number}
\eid{identifier}
\date[Date text:]{\today}
\received[Received text:]{date}

\revised[Revised text:]{date}

\accepted[Accepted text:]{date}

\published[Published text:]{date}

\startpage{1}
\endpage{12}
\maketitle
\tableofcontents

\section{Introduction}

A protein is a chain of amino acids, referred to as residues, that folds into
a characteristic shape in water at a sufficiently low temperature. The main
force comes from the hydrophobic effect, which tends to drive hydrophobic
residues to the interior of the folded conformation, or native state.
Hydrogen bonding among residues leads to the formation of alpha helices and
beta sheets. The native structure is usually described in terms of primary,
secondary, and tertiary structures, which repsectively refer to the sequence of
amino acids along the chain, the alpha helices and beta sheets mentioned, and
the gross geometrical structure\cite{BT,Daggett03}.

An interesting question is whether secondary structure emerges before
tertiary structure during folding. We try to answer the question for a protein fragment (peptide), polyalanine (Ala$_{20}$), which has 20 identical amino acids, the
hydrophobic Alanine. The native state is known to be a single alpha helix. The
tertiary structure, therefore, is a cylindrical tube.

We use the CSAW (conditioned self-avoiding walk) model propose recently by one
of us\cite{Huang07}. The idea is that the unfolded protein is a
random coil, which can be represented as a random walk that is not allowed to
cross itself. Such a self-avoiding walk (SAW) simulates the fact that
different residues cannot occupy the same location in space. The interactions
repsonsible for folding, chiefly the hydrophobic interaction and
hydrogen-bonding, are taken into by imposing conditions on the SAW, hence the
name CSAW.

In a computer simulation, one generates an ensemble of SAW and extracts a
sub-ensemble that satisfies desired conditions. The latter process is
implemented through the Monte Carlo method, which can generate a sequence of
states distributed according to a canonical ensemble. The conditions mentioned
are expressed through various energy terms in the Hamiltonian.

Mathematically, CSAW is a simulation of a Langevin equation\cite{Huang05} that
describes the folding protein chain. We refer to refs.\cite{Huang07} for details, but include a brief summary in the Appendix, and
illustrate its equivalence to molecular dynamics through a simple example.

We find that, in the case considered, the overall size of the protein has
reached an equilibrium value, while helical content continues to increase. In
this sense, large scale structure forms before small scale structure. We will be
able to see how the helix starts forming through nucleation.

The time evolution of the large and small scale structures exhibit
qualitatively different behaviors, which we can explain in terms of
phenomenological models. The formation of small-scale secondary structures are
governed by rate equations for the growth of helical content, while the
relaxation of the large-scale size is analogous to that of an elastic solid.

Since CSAW is a relatively new model, this work serves as a test of its
validity. In this respect, the model appears to be effective in describing the
dynamics of folding. When we affirm the model, we are affirming the underlying
principle as implied by the Langevin equation, namely, protein folding is a
stochastic dissipative process that tends toward thermal equilibrium with the environment.

From this point of view, the perennial debate on whether the folded state is
in therdynamic equilibrium or in some kinetic steady state is merely a
question of whether the protein can reach thermal equilibrium in realistic
time, or gets trapped in some intermediate state. The calculations here, using
experimental input, indicate that a small protein like the present one reaches
thermal equilibrium in the order of 20ns. It would be interesting to investgate
this question for large proteins.

This work is intended to address one aspect of the dynamical process of protein folding, instead of validating the model in a comprehensive way. Therefore, instead of putting everything into a comprehensive model, we focus on two most important interactions for helix formation: hydrophobic and hydrogen bonding interactions. The Monte Carlo method used is based on a simulation of the Langevin equation, which is consistent to the physical model. Despite the simplicity, this model allows us to delineate the mechanisms of the formation of different scale structures. In the simple example, we find that the evolution of small and large scale structures involve different mechanisms. This, not necessary general, can give physical insight for wider understanding of protein folding as a stochastic process.  Such physical insight will be helpful for further study, both theoretically and experimentally.

\section{Methods and results}

The initial state of the simulation was created by unfolding the native state
of Ala$_{20}$ through $4\times10^{4}$ CSAW steps, at a program temperature of
$T^{\ast}=4.4$. The value was high enough to make the protein unfold into a
random coil. We then set $T^{\ast}= 0.2$, and the folding process
began. We do not yet have a precise calibration of $T^{\ast}$ against the
physical temperature.

The folding process ran for $4\times10^{6}$ CSAW steps. The entire run was
repeated 100 times to generate an ensemble of 100 folding trajectories. Each
trajectory consumes the order of 1 hr of computer time on a workstation.

A CSAW step here means one Monte Carlo trial step, whether or not the trial
results in a successful update. It simulates real time, during which the
system tries to overcome energy barriers but does not always succeed. On
average, it takes about 30 tries to achieve an update.

According to analysis given later, which compares our results with
experimental data, our run corresponds to about 10ns in physical time\cite{Bier99,Bilsel00}. From
this we estimate that one CSAW step corresponds to approximately $10^{-15}$ s.

During folding, the helical content rises from an average initial value of
0.05 to 0.55, and tends toward an asymptote of 0.77. This
indicates that we have not reached the native state. The ensemble generated is
an evolving ensemble that is not yet canonical. This suits our purpose, which
is to study the folding dynamics. Various relaxation times can be obtained by
analyzing the evolution towards equilibrium.

We follow the evolution at different length scales by measuring the average
radius of gyration $R_{g}$, average length of helix segments $L_{H}$, and
average helicity $f_{H}$. They are defined by%
\begin{align}
R_g^{2}(t)  &  =\frac{1}{2N^2}\sum_{n,m=1}^{N}\left\langle \left(  \mathbf{R}%
_{n}(t)-\mathbf{R}_{m}(t)\right)  ^{2}\right\rangle ,\nonumber\\
L_{H}(t)  &  =\left\langle \text{Av. length of helix segment}\right\rangle
,\nonumber\\
f_{H}(t)  &  =\left\langle \frac{\text{No. residues in helix}}{\text{Maximum
No.}}\right\rangle ,
\end{align}
where $N$ is the number of residues, $\mathbf{R}_{n}(t)$ is the instantaneous position
of the center of the $n$th residue, and $\left\langle {}\right\rangle $
denotes ensemble average at time $t$. These quantities respectively measure
structures on the largest, intermediate, and smallest length scales.

We also measure the structure factor%
\begin{equation}
g(k,t)=\frac{1}{N}\sum_{n,m=1}^{N}\left\langle \exp\left(  i\mathbf{k}%
\cdot\left[  \mathbf{R}_{n}(t)-\mathbf{R}_{m}(t)\right]  \right)
\right\rangle ,
\end{equation}
which is the Fourier transform of the density correlation function, accessible
to experiments through x-ray scattering. It is independent of the directions
of $\mathbf{k}$ because of the ensemble average\cite{Doi01}.

\begin{figure}[h]
 \centering
\includegraphics[width=12cm]{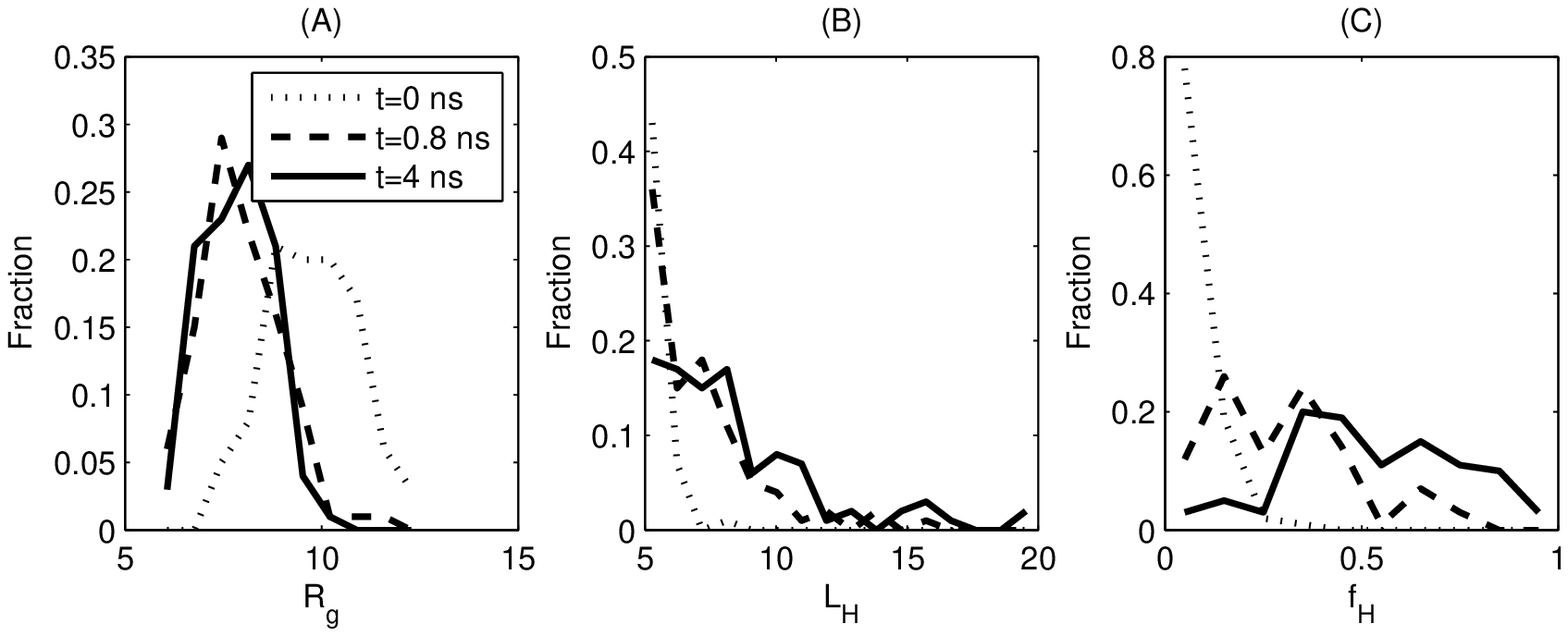}
\caption{}
\label{fig:1}
\end{figure}

Fig.\ref{fig:1} shows distributions of $R_{g}$, $L_{H},$ $f_{H}$ at different times.  Fig.\ref{fig:2} shows $R_{g}$, $L_{H},$ $f_{H}$ as functions of time on a log scale.
Lines through the calculated points are fit made to obtain relaxation times,
to be detailed later.

The behavior of \ $R_{g}$ shows that there was a very fast collapse, followed
by two slower stages. Such a two-stage behavior has been observed
experimentally in larger proteins\cite{Akiyama02,Uzawa04}. We shall
analyze them later in terms of theoretical models.

As we can see from Fig.\ref{fig:2}, there is little change in $R_{g}$ apart from
fluctuations after 1ns, but $L_{H}$ and $f_{H}$ continue to increase.
This means that, while the overall size of the protein has equilibriated, the
secondary structures continue to adjust. We shall quantify this in terms of
relaxation times.

Fig.\ref{fig:3} shows $g(k)$ as functions of $k$, for different times (see caption for detail).

Fig.\ref{fig:4} shows a contour map of the ensemble average of local helicity. The
vertical axis is residue number, and the horizontal axis is time. Helix
nucleation started near residues 6 and 14. Since the contour plot here is an
ensemble average, this indicated that the nucleation points are not random,
but occur at specific positions, at least for this small protein.

\section{Analysis and discussion}

\begin{figure}[h]
 \centering
\includegraphics[width=12cm]{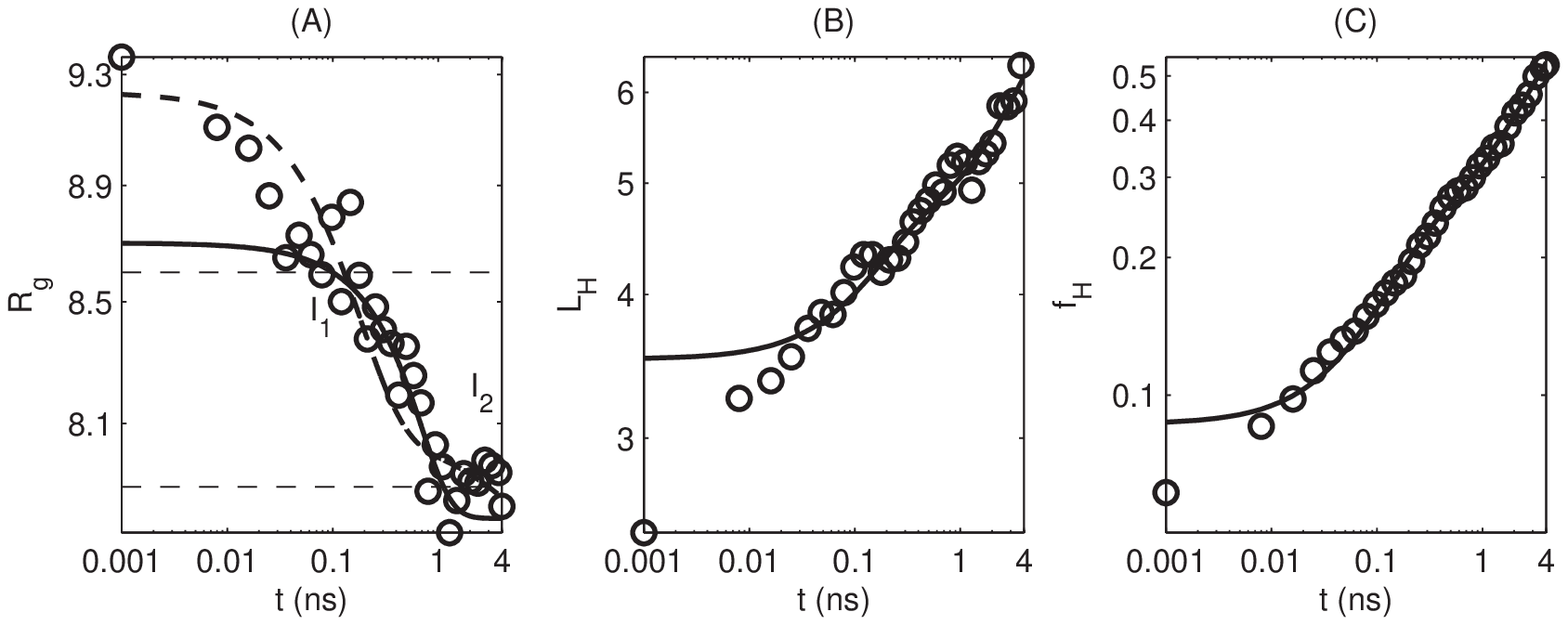}
\caption{}
\label{fig:2}
\end{figure}

\begin{figure}[h]
 \centering
\includegraphics[width=12cm]{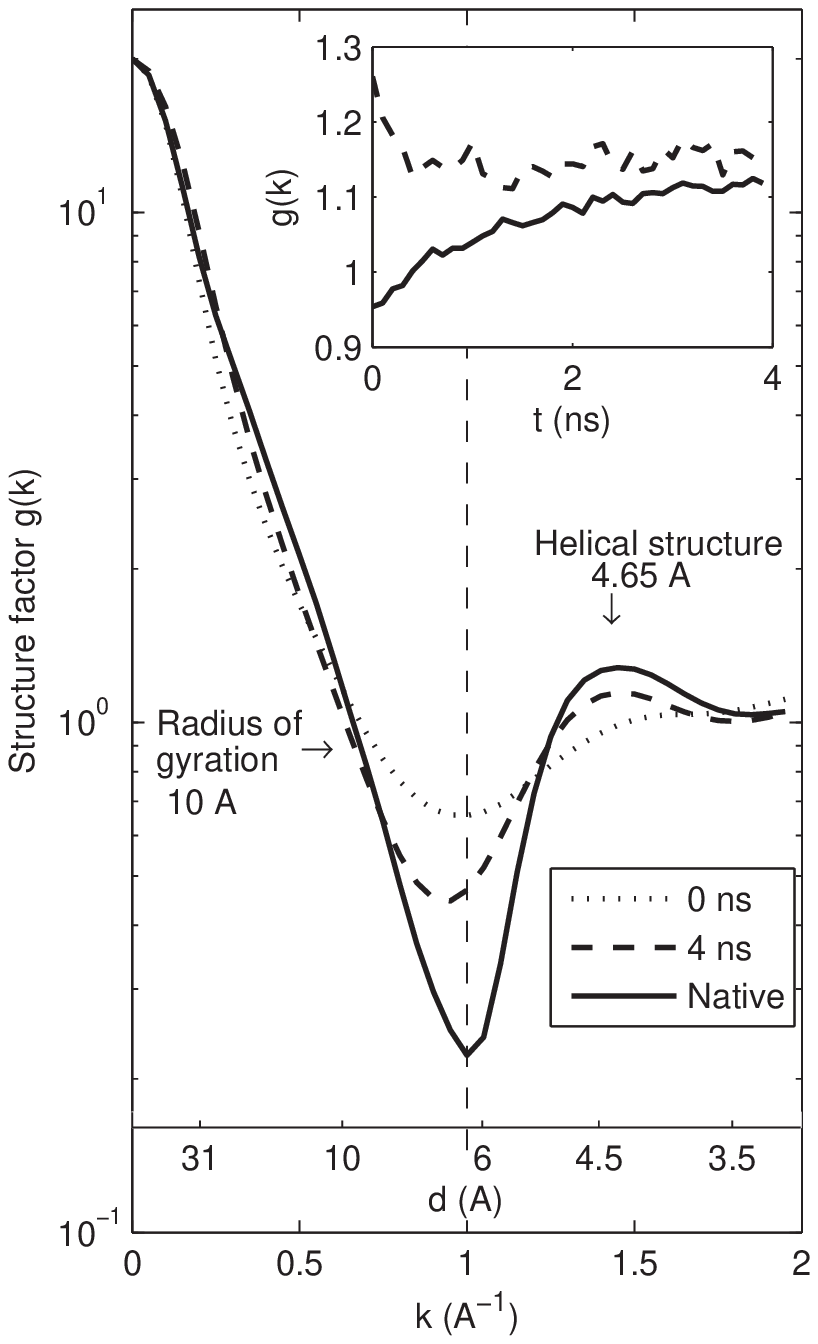}
\caption{}
\label{fig:3}
\end{figure}

\begin{figure}[h]
 \centering
\includegraphics[width=12cm]{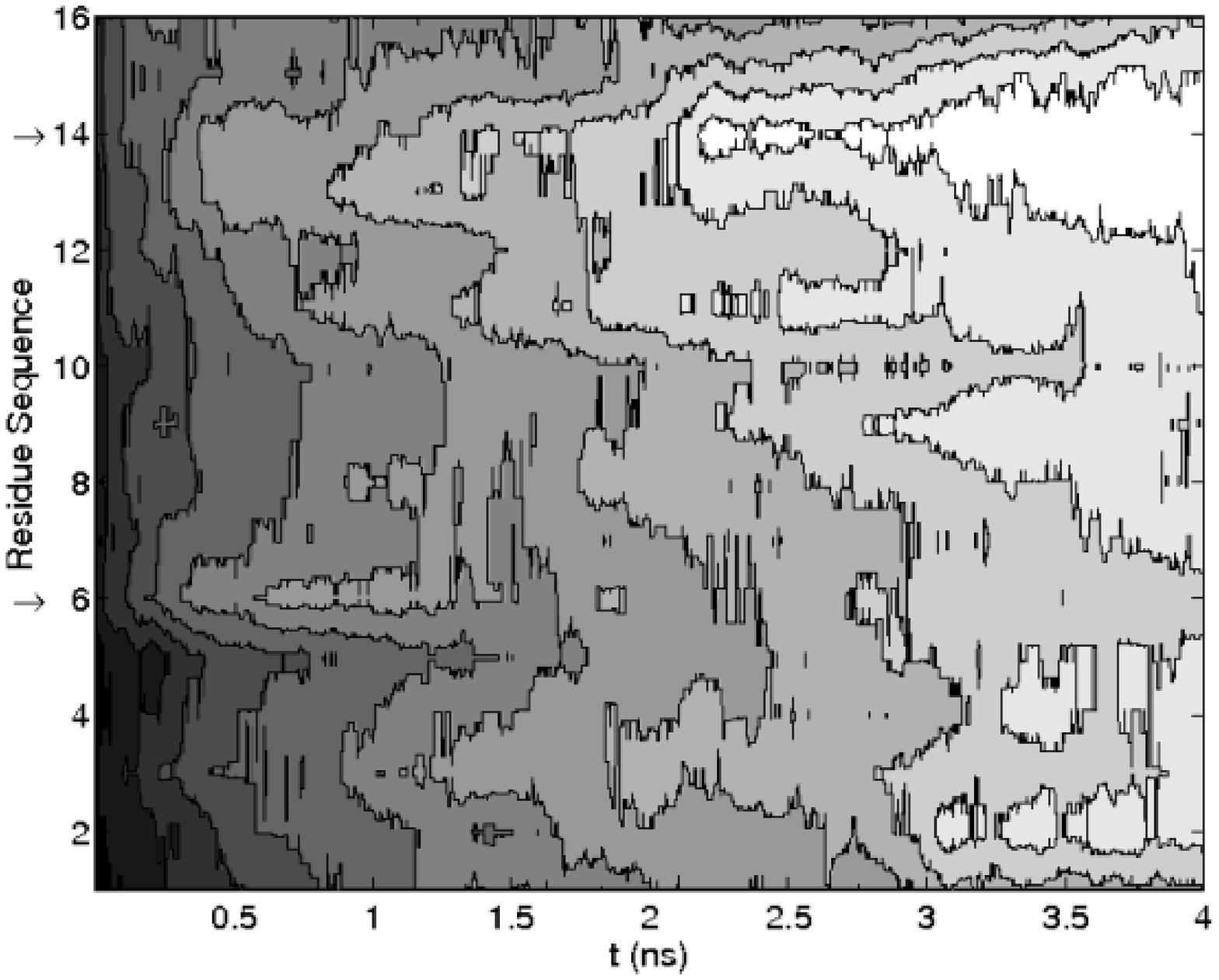}
\caption{}
\label{fig:4}
\end{figure}

The existence of two folding stages suggest that we fit the late-time evolutions with two exponential functions. Indeed, we obtain good fits
for $L_{H}$,$f_{H}$ with
\begin{equation}
 \begin{array}{rcl}
L_{H}(t)  &  =& 7.43-0.98e^{-t/0.17}-2.95e^{-t/4.68},\\
f_{H}(t)  &  =& 0.77-0.13e^{-t/0.17}-0.55e^{-t/4.68}.
\end{array}
\end{equation}
where the unit for $t$ is ns. These are shown as solid curves in Fig.\ref{fig:2}. They
suggest that the ensemble will reach equilibrium at 20ns, with average helical
content $0.77$. The relaxation times for the two stages are $0.17$ns and
4.68ns, respectively.

The time scale is determined as follows. Originally $t$ was measured in CSAW
steps. We judge that at the end of our runs the folding process was about 70\%
complete, and that puts the halfway point at about 3$\times10^{6}$ steps.
Identifying this with the experimental value of $t_{1/2}=16$ns \cite{Bier99,Bilsel00}, we
arrive at the estimate of approximately $10^{-15}$s per step.

\begin{figure}[h]
 \centering
\includegraphics[width=12cm]{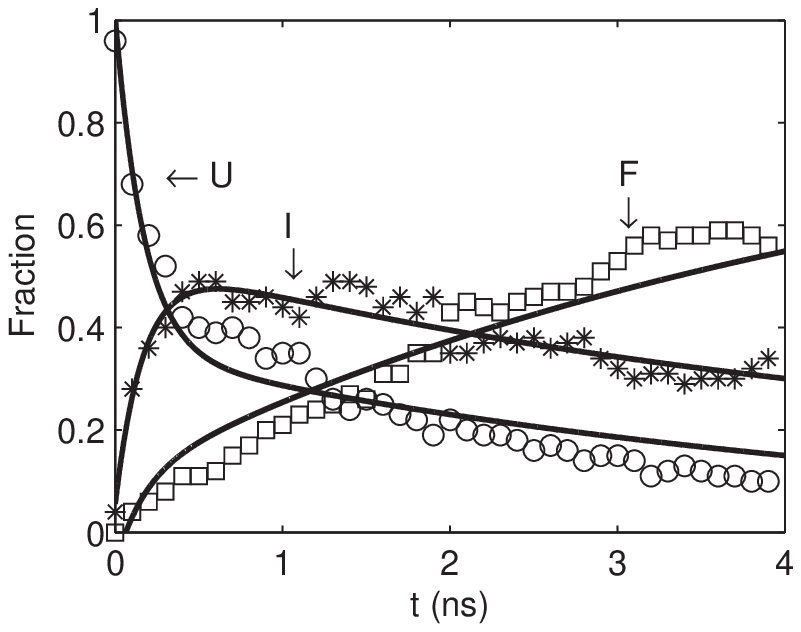}
\caption{}
\label{fig:5}
\end{figure}

The two-stage behavior of the development of secondary structure suggests the
following model. We picture the ensemble as a mixture of three classes of
protein chains: unfolded (U) with $f_{H}<02,$ intermediate (I) with
$0.2<f_{H}<0.5$, and folded (F) with $f_{H}>0.5$. There are three-state
transitions among these classes:%
\begin{equation}
\text{U}\rightleftharpoons\,\text{I}\rightleftharpoons\,\text{F\thinspace}.
\end{equation}
The relative fractions of these classes evolve with time. These fractions can be obtained by
solving rate equations, using time constants given prevously. They can of
course be extracted from our simulation data. Fig.\ref{fig:5} shows that the two agree
rather well.

The two-exponential fit does not work for $R_{g}$, as we see by the dashed
curve in Fig.\ref{fig:2}. Thus, the relaxation of of $R_{g}$ calls for a different
mechanism. For this, we model the gross structure as an elastic solid, with an
effective potential energy%
\begin{equation}
V(R_{g})=\left(  \frac{A}{R_{g}}\right)  ^{11}-\left(  \frac{B}{R_{g}}\right)
^{5},
\end{equation}
and a phenomenological equation of motion%
\begin{equation}
\gamma \frac{dR_{g}}{dt}=-\frac{dV}{dR_{g}},
\end{equation}
where $t$ is in ns, $R_{g}$ in A. Solving the equation with $\gamma = 3.33, A=9.58,
B=14.35$, initial condition $R_{g}(0)=8.7$, we obtain the solid curve in
Fig.\ref{fig:2}, which gives a good fit to the simulation data.

Fig.\ref{fig:2} shows that $R_{g}$ reaches equilibrum after about 1 ns, at a value
slightly lower than the data points. This suggests that there are
perturbations to radius relaxation from the secondary structure, which
continues to undergo adjustment.

\section{Relation to other works}

\begin{figure}[h]
 \centering
\includegraphics[width=12cm]{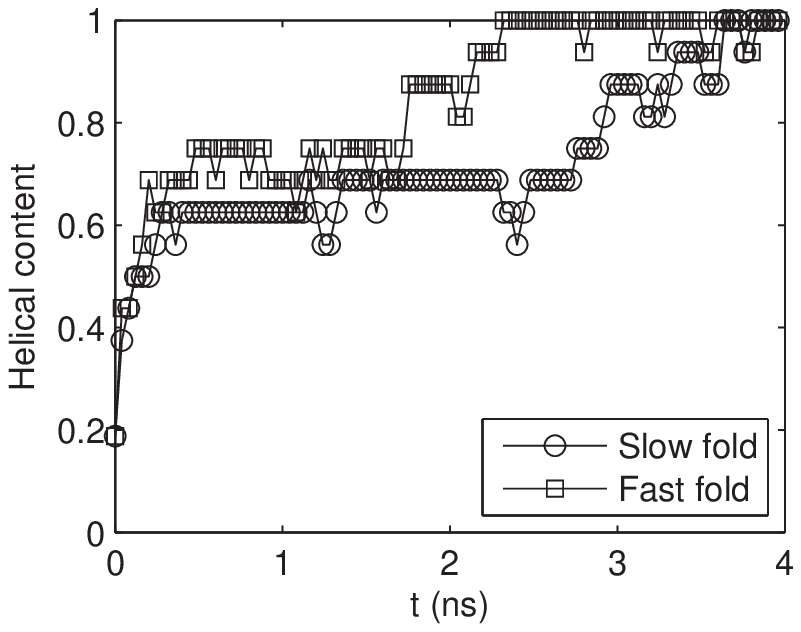}
\caption{}
\label{fig:6}
\end{figure}

In our ensemble, we find that there are two types of evolution paths, a fast
and a slow one, as illustrated in Fig.\ref{fig:6}. This supports results from discontinuous molecular
dyanmics\cite{Smith98}. We have found the same fast and slow paths in the
folding of chignolin, a 10-residue synthetic peptide\cite{Lei}. The meaning of
this is not yet clear.

\begin{figure}[h]
 \centering
\includegraphics[width=12cm]{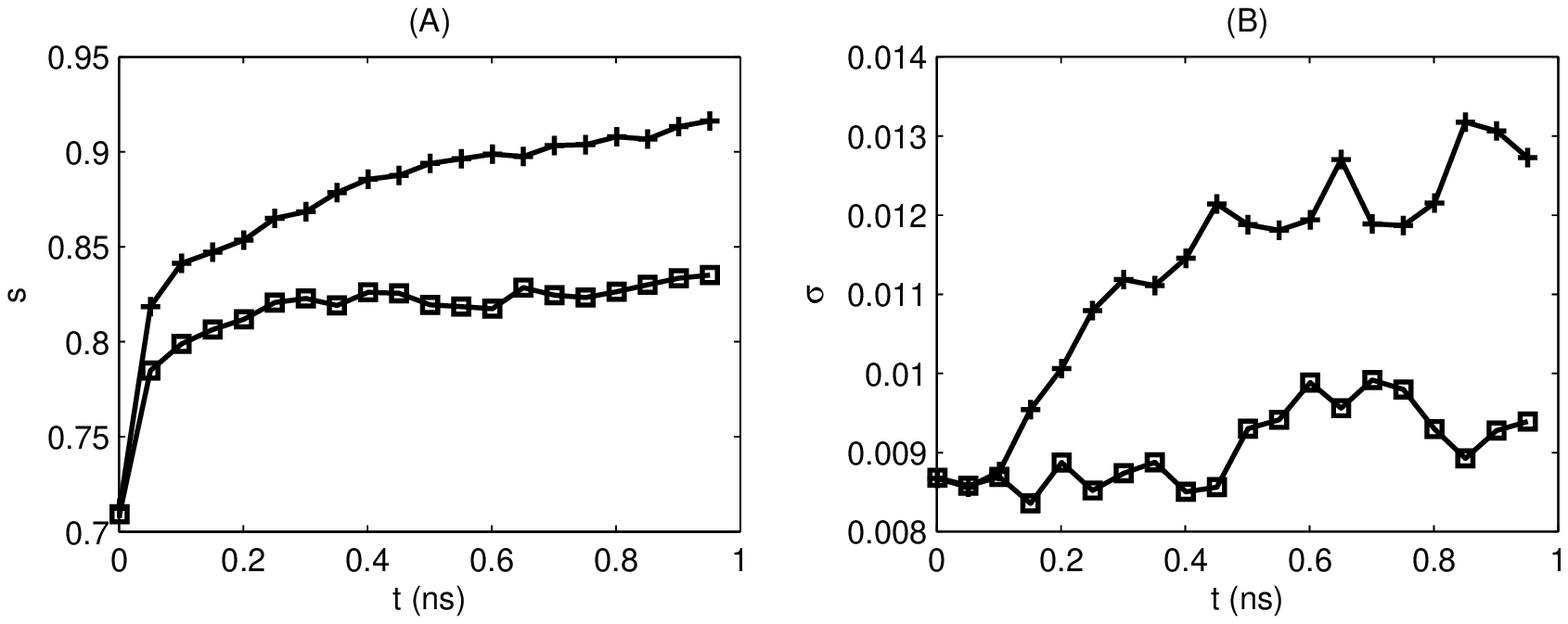}
\caption{}
\label{fig:7}
\end{figure}

In an early work on alpha-helix formation, Zimm and Bragg introduced
parameters $s$ and $\sigma$, which respectively measures the probability of
helix growth and nucleation. The quantities of these two parameters  are related to $f_H$ and $L_H$ according to following equations\cite{Shen},
\begin{eqnarray}
f_H &=& \dfrac{1}{2}-\dfrac{1-s}{2\sqrt{(1-s)^2+4 s \sigma}}\\
L_H &=&1 +\dfrac{2 s}{1-s+\sqrt{(1-s)^2+4 s \sigma}}.
\end{eqnarray}
We have calculated these quantities using CSAW. 
An advantage in our model is that we can turn on or off selected interactions. In
Fig.\ref{fig:7} we show the results with and without the hydrophobic effect. We can see
that the hydrophobic effect has pronounced influence on helix nucleation, but
it is not as important for helix growth.

\appendix

\section{CSAW (Conditioned Self-Avoiding Walk)}

The CSAW model is an algorithm that successively updates a protein state, in
order to generate a canonical ensemble of states. Starting with an initial
state which is an arbitrary non-overlapping chain (SAW), we generate a new
chain by the pivoting algorithm, and keep doing so until we obtain another
non-overlapping chain (a new SAW). We then decide whether to accept this as
an update via the Metropolis Monte Carlo method, as follows. We ask whether the
proposed update decreases the energy $E$. If it does we accept it, and
otherwise accept it with a relative probability given by the Boltzmann factor%
\[
p=\exp\left(  -\left(  E_{\text{new}}-E_{\text{old}}\right)  /k_{B}T\right)
.
\]
That the energy can increase simulates thermal fluctuations, and makes the
updating process one of minimizing the free energy.

Along the backbone of the protein chain a series of carbon atoms (the
$C_{\alpha})$ are connected by covalent chemical bonds that are shaped like a
crank, which lies in one plane. The major degrees of freedom of the chain are
the torsional angles that define the relative orientation of two successive
planes. Other degrees of freedom, such as small vibrations of the chemical
bonds, can be neglected when we consider protein folding. The state of a
protein of $N$ residues are thus specified by $N-1$ pairs of torsional angles.
These are the only degrees of freedom considered.

For the present study, side chains are approximated by hard spheres, and
other atoms are treated as hard spheres with known van der Waals radii. The
only interactions included are those corresponding to the hydrophobic effect,
and hydrogen bonding. The energy is taken to be%
\begin{equation}
E=-g_{1}K_{1}-g_{2}K_{2},
\end{equation}
where: $K_{1}$ is the total hydorphobic contact number, i.e., the total number of
nearest neighbors surrounding a hydrophobic residue, (not counting the two
permanent nearest neighbors along the chain.) In the present case, all residues
are hydrophobic. The quantity $K_{2}$ is the total number of internal hydrogen
bonds, which connects hydrogen to oxygen in different residues. Such a bond is
deemed to exist whenever the partner are within a certain range of distrance
from each other, and the chemical bonds they are attached to are antiparallel
within given margins.

In general, the clear separation of hydrophobic effect and hydrogen bonding is
an approximation, under the assumption that atoms on the backbone can only
bond with one another, while those on the side chains can only bond with
water. In the present case this distinction is moot, since all rersidues are
hydrophobic, whose side chains cannot form hydrogen bonds.

\begin{figure}[h]
 \centering
\includegraphics[width=12cm]{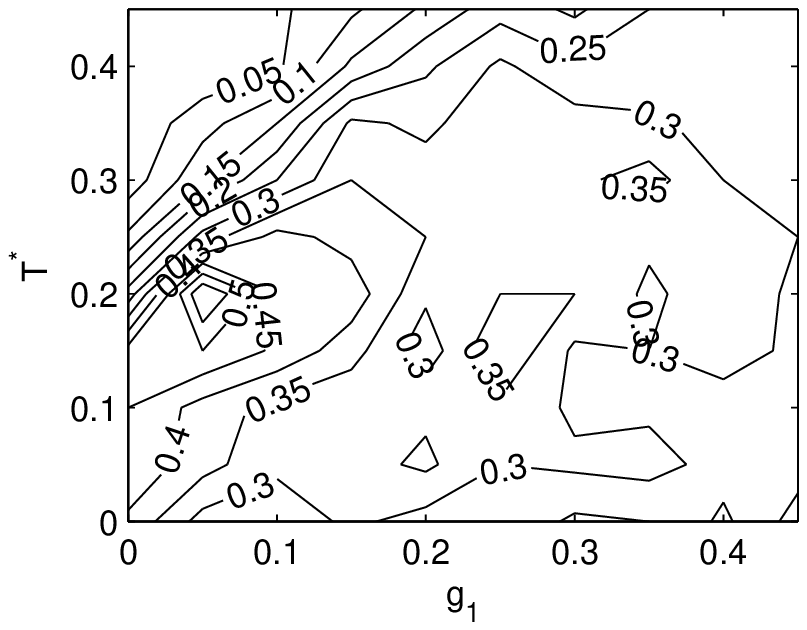}
\caption{}
\label{fig:8}
\end{figure}

There are two independent parameters $g_{1}$ and $g_{2}$ in the model.
Actually, in the Monte Carlo procedure, only the combinations $g_{1}/k_{B}T$
and $g_{2}/k_{B}T$ are relevant. We define a program temperature $T^{\ast
}=k_{B}T/g_{2},$ and use $T^{\ast}$ and $g_{1}/g_{2}$ as independent parameters. To simplify the notation, we set $g_{2}=1$. To fix the parameters, we
calculate the helical content $f_{H}$ for various values, and choose that which
gives the maximum helicity after 10$^{6}$ CSAW steps. The contour map of
$f_{H}$ so obtained is shown in Fig.\ref{fig:8}, from which we pick the values
$g_{1}=0.05,$ $T^{\ast}=0.2$.

The CSAW model is a computer simulation of a generalized Langevin equation for
the protein chain, which has the form%
\begin{equation}
m_{k}\mathbf{\ddot{r}}_{k}=-m_{k}\gamma_{k}\mathbf{\dot{r}}_{k}+\mathbf{F}%
_{k}(t)+\mathbf{G}_k(\mathbf{r}_{1},\cdots,\mathbf{r}_{N}),\ \ \ (k=1,\cdots,N)
\end{equation}
where $\mathbf{r}_{i}$ is the position of the $i$th atom, $m_{k}$ its mass,
$\gamma_{k}$ its dissipation coefficient, and $\mathbf{F}_{k}$ the random
force acting on it by the medium. The term $\mathbf{G}_{k}$ is a symbolic
representation of all the non-random forces acting on the atom, including
interaction with other atoms in the protein, the chemical bonds that hold the
chain together, and the hydrophobic interaction with the medium. It is highly
improbable that we can solve this equation analytically, but we can simulate
on a computer. The dissipation and random force is simulated by random walk,
and the forces in $\mathbf{G}_{k}$ that maintain the chain and prevent atoms
from overlapping make the random walk a SAW. The rest of the forces in
$\mathbf{G}_{k}$ is taken into account through Monte Carlo.

\begin{figure}[h]
 \centering
\includegraphics[width=12cm]{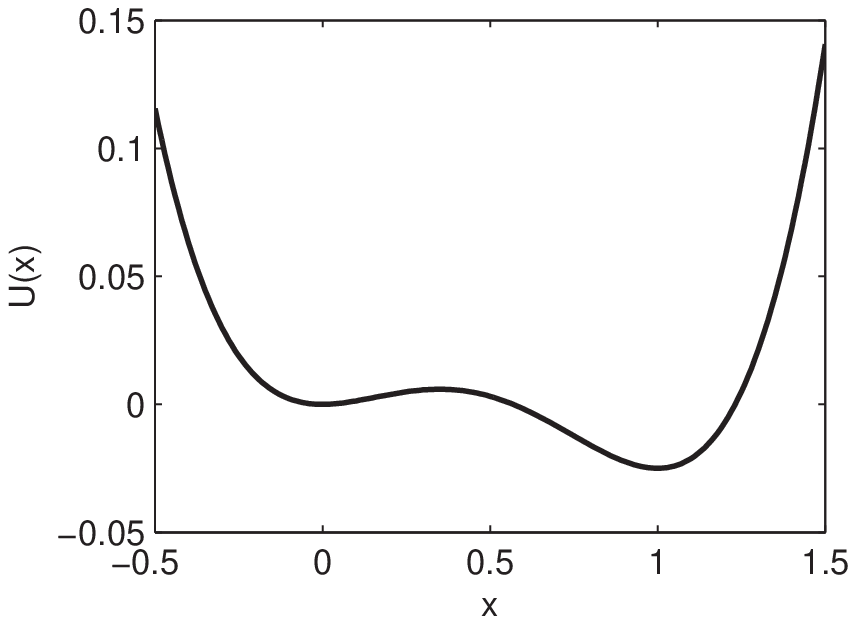}
\caption{}
\label{fig:9}
\end{figure}

\begin{figure}[h]
 \centering
\includegraphics[width=12cm]{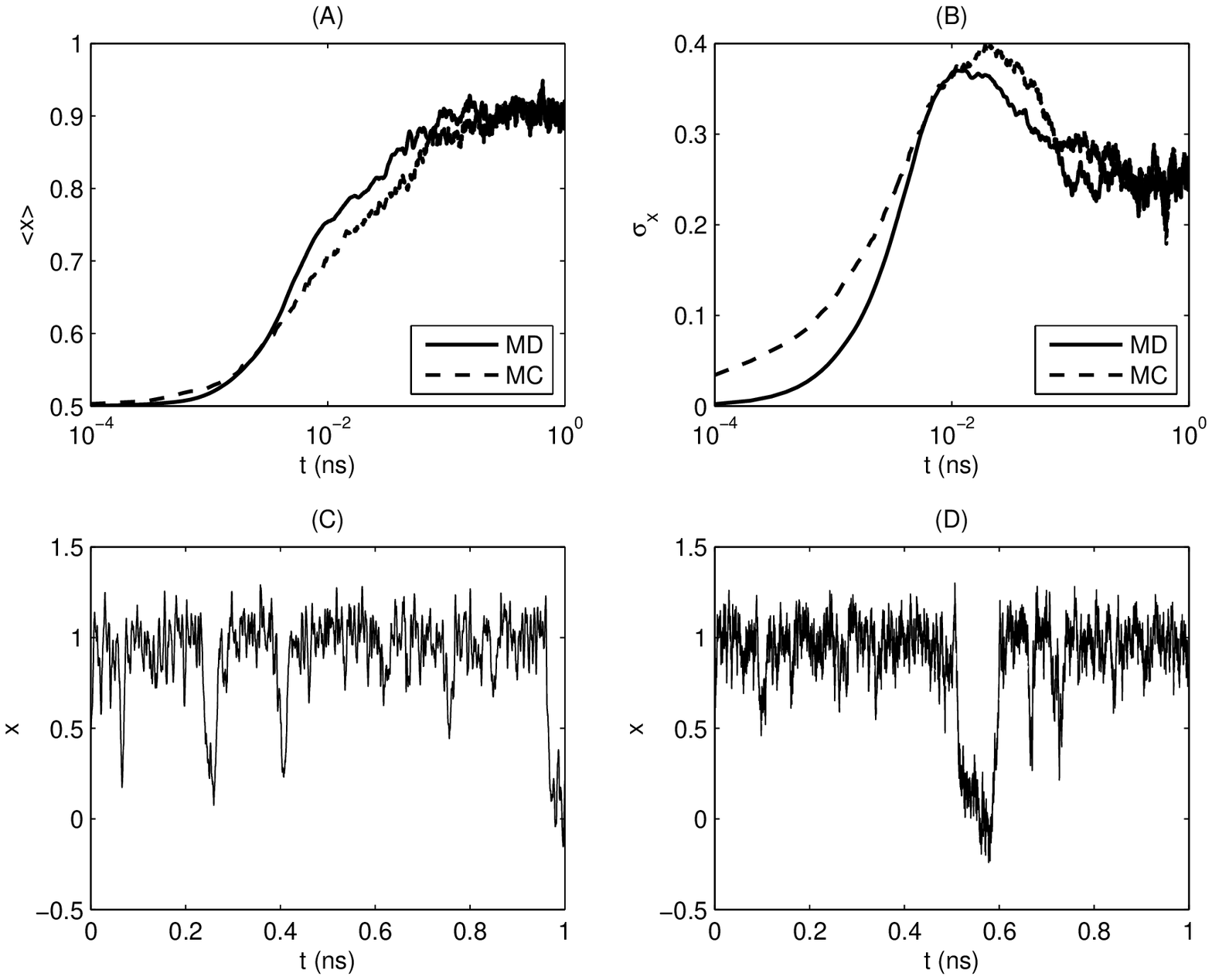}
\caption{}
\label{fig:10}
\end{figure}

To illustrate that this procedure yields a solution of the equation, we
consider a simpler case, the Brownian motion of a particle in 1D, in a
potential well. The Lagevin equation reads%
\begin{equation}
\ddot{x}=-\frac{dU(x)}{dx}-\gamma\dot{x}+F(t),
\end{equation}
with a double-well potential
\begin{equation}
U(x)=a\left(  \frac{1}{2}x^{2}-\frac{1}{3}x^{3}\right)  +\left(  \frac{1}%
{4}x^{4}-\frac{1}{3}x^{3}\right)  ,
\end{equation}
which is sketched in Fig.\ref{fig:9}, with $a=3.5.$ The equation is then solved as a
stochastic differentiation using molecular dynamics, and alternatively using
Monte Carlo. Comparison of these two methods are given in Fig.\ref{fig:10}, which show the
equivalence in a statistical sense.

\newpage

\section*{Figure Captions}

Fig.1 \ Distribution function for (A) radius of gyration $R_{g},$ (B) average
length of helix segment $L_{H}$, (C) local helicity $f_{H\text{ \ }}$. They
pertain to structures on large, intermediate and small length scales,
respectively.\ The distributions are shown for three different times during
folding, which lasts 4 ns.

Fig.2. Time evolution of (A) radius of gyration $R_{g},$ (B) average length of
helix segment $L_{H}$, (C) local helicity $f_{H}$. For (B) and (C), the solid
curves are fits by a sum of two exponentials, which can be derived from rate
equations governing the growth of helicity. The two-exponential for (A), shown
as the dashed curve, is not satisfactory. Instead, a better fit (solid curve)
is obtained via a model that treats the protein as an elastic solid. This
shows that the relaxation of large and small scale structures are governed by
different mechanisms.

\bigskip

Fig.3. The structure function $g(k)$ is the Fourier tranform of the density
correlation function, and contains information about structures on different
length scales. It is shown at different times during the folding process. The inset show the time evolution at two specific wave numbers $k$, corresponding respectively to large (dashed curve) and small (solid curve) scale structures.

Fig.4. Contour map of ensemble average of local helicity. Vertical axis is
residue sequence along the protein chain, and horizontal axis is time. Arrows
point to points of nucleation of helical structure. 

\bigskip

Fig.5. The ensemble can be divided into classes of protein chains
characterized by different helical content: U--unfolded, I--intermediate,
F--folded. Data points are from CSAW simulation, and solid curves are
calculated from phenomenological rate equations governing transitions among
these classes. The rate equations underlie the two-exponental fits in (B) and
(C) of Fig.\ref{fig:2}.

\bigskip

Fig.6. The evolution of helicity reveals two types of folding paths,
fast and slow. This reproduces results of other works, but its significance is
yet to be understood.

\bigskip

Fig.7. Evolution of the probability of helix growth (s), and helix nucleation
($\sigma).$ The hydorphobic effect is turned on in the upper curves, and off
in the lower cureves. It has a more prounced effect on nucleation than on growth.

\bigskip

Fig.8. Contour map of average helicity as functions of CSAW parameters.
We choose the set of parameters at the maximum helicity.

\bigskip

Fig.9. Double-well potential used in the Lagevin equation in illustrative calculations.

\bigskip

Fig.10. Demonstration of the statistical equivalence of MC (Monte Carlo) and  MD
(molecular dynamics) in solving the illustrative Langevin equation. (A) and
(B) show the average position and standard deviation as functions of time. (C) and (D) show
sample paths from MC and MD respectively. We can see that the position $x$ go over
the energy barrier to visit the origin, but at different times.

\newpage

\section*{Figures}

\end{document}